\documentclass[onecolumn,showpacs]{revtex4}

\usepackage{graphicx}% Include figure files
\usepackage{dcolumn}% Align table columns on decimal point
\usepackage{bm}% bold math

\input epsf

\begin{document}

\title{\Large  Interaction between phantom field and modified Chaplygin gas}

\author{\bf Surajit
Chattopadhyay$^1$\footnote{surajit$_{-}$2008@yahoo.co.in} and
~Ujjal Debnath$^2$\footnote{ujjaldebnath@yahoo.com}}

\affiliation{ $^1$Department of Computer Application, Pailan
College of Management and Technology, Kolkata-104, India.\\
$^2$Department of Mathematics, Bengal engineering and Science
University, Howrah-103, India.}

\date{\today}

\begin{abstract}
In this letter, we have considered a flat FRW universe. Instead of
considering only one candidate for dark energy, we have considered
interaction between phantom field and modified Chaplygin gas. It
has been shown that the potential of the phantom field increases
from a lower value with evolution of the Universe. It has been
observed that, the field has an increasing tendency and potential
has also an increasing tendency with passage of cosmic time. In
the evolution of the universe the crossing of $w=-1$ has been
realized by this interacting model.\\
\end{abstract}


\maketitle

In recent observations it has been established that the Universe
 is made up of roughly two-thirds dark energy
that has a negative pressure and can drive the accelerating
expansion of the Universe and a plethora of literatures have
described these facts [1]. Numerous models have been proposed as
candidates of dark energy. Examples of such candidates for dark
energy are vacuum energy, quintessence [2], phantom [3] and
Chaplygian gas [4]. In order to derive the accelerated expansion
of the Universe it is required to have $w<-\frac{1}{3}$. For
cosmological constant $w=-1$, for quintessence model
$-1<w<-\frac{1}{3}$ and for k-essence model $w<-1$ or $w>-1$.
However, it is physically impossible to cross $w=-1$ [5]. Present
observation data constrain the the range of the equation of state
of dark energy as $-1.38<w<-0.82$ [6]. However, the equation of
state of conventional quintessence models that are based on a
scalar field and positive kinetic energy cannot evolve to the the
regime of $w<-1$. Some authors have investigated a phantom field
model which has a
negative kinetic energy and can realize $w<-1$ in its evolution [7]. \\

Observations have shown that the current universe is very close to
a spatially flat geometry [8]. This is actually a natural result
from inflation in the early universe [9]. Consequently, in many
recent works the Universe is considered to be spatially flat $(k =
0)$ [10 and references therein]. However, recent observations have
shown that the spatial curvature of the universe is small.
Nevertheless, this result is not sufficient to set conclusions
about the topology of the universe. Even in the case of very small
curvature, a non-flat model could imply in important consequences
on the evolution of the Universe [11].\\

 Neither dark matter nor dark energy has laboratory
evidence for its existence directly. Recently, it has been
suggested that the change of behavior of the missing energy
density might be regulated by the change in the equation of state
of the background fluid instead of the form of the potential. This
is done by introducing an exotic background fluid, Chaplygin gas,
within the framework of Friedman-Robertson-Walker (FRW) cosmology.
The Chaplygin gas is characterized by an exotic equation of state
$p=-\frac{B}{\rho}$, where $B$ is a positive constant [12]. Role
of Chaplygin gas in the accelerated universe has been studied by
several authors. The above mentioned equation of state has been
generalized to $p=-\frac{B}{\rho^{\alpha}}$ with $0\leq\alpha\leq
1$. This is called generalized Chaplygin gas [13]. This equation
has been further modified to $p=A\rho-\frac{B}{\rho^{\alpha}}$
with $0\leq\alpha\leq 1$. This is called modified Chaplygin gas
[14]. This equation of state shows radiation era at one extreme
and $\Lambda CDM$ model at the other extreme.\\

To obtain a suitable evolution of the Universe an interaction is
often assumed such that the decay rate should be proportional to
the present value of the Hubble parameter for good fit to the
expansion history of the Universe as determined by the Supernovae
and CMB data [15]. Guo and Zhang [16] investigated an interaction
between matter fluid and phantom field. In the present paper, we
have considered an interaction between phantom field and variable
Chaplygin gas. The endeavour is to consider a model comprising of
two component mixture and to discern how the interaction between
two such dark energies influence the evolution and total life time
of the universe. To do the same, we have introduced an interaction
term which is proportional to the Hubble parameter
$H=\frac{\dot{a}}{a}$ and the
density of the variable modified Chaplygin gas. \\

The metric for a spatially flat isotropic and homogeneous FRW
Universe is given by

\begin{equation}
ds^{2}=dt^{2}-a^{2}(t)\left[dr^{2}+r^{2}(d\theta^{2}+\sin^{2}\theta
d\phi^{2})\right]
\end{equation}

where, $a(t)$ is the scale factor.\\

The Einstein field equations are

\begin{equation}
3H^{2}=\rho_{tot}
\end{equation}
and
\begin{equation}
6(\dot{H}+H^{2})=-(\rho_{tot}+3p_{tot})
\end{equation}

where $\rho_{tot}$ and $p_{tot}$ are the total energy density and
isotropic pressure respectively (choosing $8\pi G=c=1$).\\

The energy conservation equation is

\begin{equation}
\dot{\rho}_{tot}+3H(\rho_{tot}+p_{tot})=0
\end{equation}

Considering a two-component model we can write

\begin{equation}
\rho_{tot}=\rho_{1}+\rho_{2}
\end{equation}
and
\begin{equation}
p_{tot}=p_{1}+p_{2}
\end{equation}

where, $\rho_{1},~\rho_{2}$ and $p_{1},~p_{2}$ are energy
densities and pressures of phantom field and modified Chaplygin
gas respectively.\\

The stress energy tensor for the phantom field is given by [17]

\begin{equation}
T_{\mu\nu}^{ph}=-\partial_{\mu}\phi\partial_{\nu}\phi+
g_{\mu\nu}\left[\frac{1}{2}g^{\alpha\beta}\partial_{\alpha}\phi\partial_{\beta}\phi-V(\phi)\right]
\end{equation}

Under the assumption that the phantom field is evolving in an
isotropic and homogeneous space-time and that $\phi$ is a function
of time alone the energy density $\rho_{1}$ and pressure $p_{1}$
obtained from $T_{\mu\nu}^{ph}$ are [17]

\begin{equation}
\begin{array}{l}
\rho_{1}=-\frac{1}{2}\dot{\phi}^{2}+V(\phi)\\\\
p_{1}=-\frac{1}{2}\dot{\phi}^{2}-V(\phi)
\end{array}
\end{equation}

Thus, the equation of state parameter is now given by

\begin{equation}
w_{1}=\frac{\frac{1}{2}\dot{\phi}^{2}+V(\phi)}{\frac{1}{2}\dot{\phi}^{2}-V(\phi)}
\end{equation}

From the above expression we find that the equation of state
parameter for phantom model is always $<-1$.\\

For modified Chaplygin gas the equation of state is given by

\begin{equation}
p_{2}=A\rho_{2}-\frac{B}{\rho_{2}^{\alpha}}
\end{equation}

where, $0\leq\alpha\leq1$ and $A$ and $B$ are positive constants.\\

Now, consider that the phantom field interacting with Chaplygin
gas, so the continuity equations for these two fluids can be
written through the interaction term $Q$ [18]

\begin{equation}
\begin{array}{l}
\dot{\rho}_{1}+3H(\rho_{1}+p_{1})=Q\\\\
\dot{\rho}_{2}+3H(\rho_{2}+p_{2})=-Q
\end{array}
\end{equation}

Note that if $Q>0$ we have that there exists a transfer of energy
from the fluid $\rho_{2}$ to the $\rho_{1}$. If $Q=0$ then we get
the non-interacting situation. The nature of the $Q$ term is not
clear at all. It may arise in principle from some microscopic
mechanisms [18]. For solving these equations different forms for
the interaction term $Q$ have been considered in the literature.
In the present paper we have taken $Q$ as $Q=3\delta H\rho_{2}$
[19], where $\delta$ is the coupling constant. Solving the second
continuity equation we get

\begin{equation}
\rho_{2}=\left(\frac{B}{1+A+\delta}+\frac{C}{a^{3(1+A+\delta)(1+\alpha)}}\right)^{\frac{1}{1+\alpha}}
\end{equation}

where, $C$ is the constant of integration.  If $A$ and $B$ tend to
zero, then $\rho_{2}\sim a^{-3(1+\delta)}$. In this case, if we
choose $\delta=0$ then Chaplygin gas behaves like dust matter and
$\delta=1/3$ gives radiation dominated universe. Now for
simplicity, we choose $V=n\dot{\phi}^{2}$, where $n$ is a positive
constant. From equation (8), it is observed that $n$ should be
greater than $\frac{1}{2}$. Using the above solution for
$\rho_{2}$ in the first continuity equation, we obtain

\begin{equation}
\dot{\phi}^{2}=-\frac{B\delta}{1+A+\delta}+\frac{6C\delta
a^{-3(1+\alpha)(1+A+\delta)}}{(3-6n)(1+\alpha)(1+A+\delta)-6}
+C_{1}a^{\frac{6}{2n-1}}
\end{equation}

and

\begin{equation}
V=nC_{1}a^{\frac{6}{2n-1}}-n\delta\left(\frac{B}{1+A+\delta}+\frac{Ca^{-3(1+\alpha)(1+A+\delta)}}{1+\frac{1}{2}(2n-1)(1+\alpha)(1+A+\delta)}\right)
\end{equation}

\begin{figure}
\includegraphics[height=1.8in]{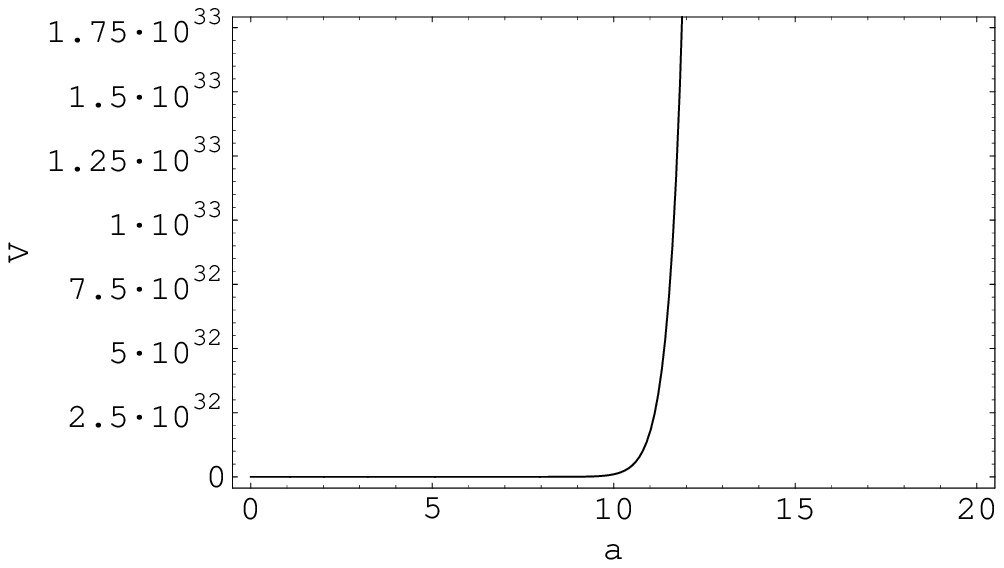}\\
\vspace{1mm} ~~~~~~~~~~~~Fig.1~~~~~~~~~~~~~~~~~\\

\vspace{7mm}

Fig. 1 represents the variation of $V$ against $a$ for
$A=0.1,B=0.1,C=0.001,\delta=0.005,\alpha=0.5,n=0.6$.

\vspace{6mm}

\end{figure}

\begin{figure}
\includegraphics[height=1.8in]{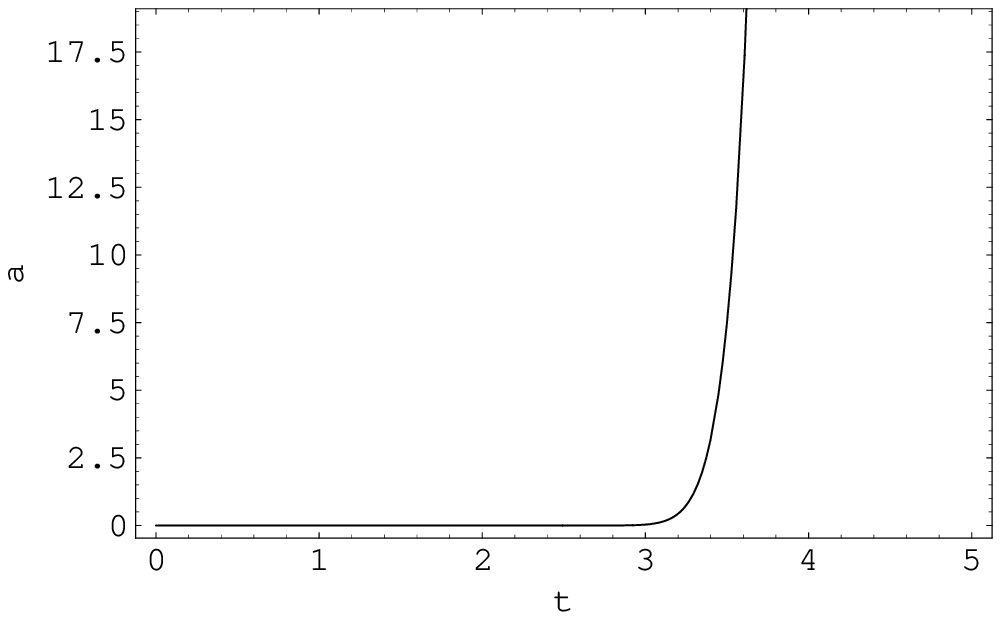}~~~~
\includegraphics[height=1.8in]{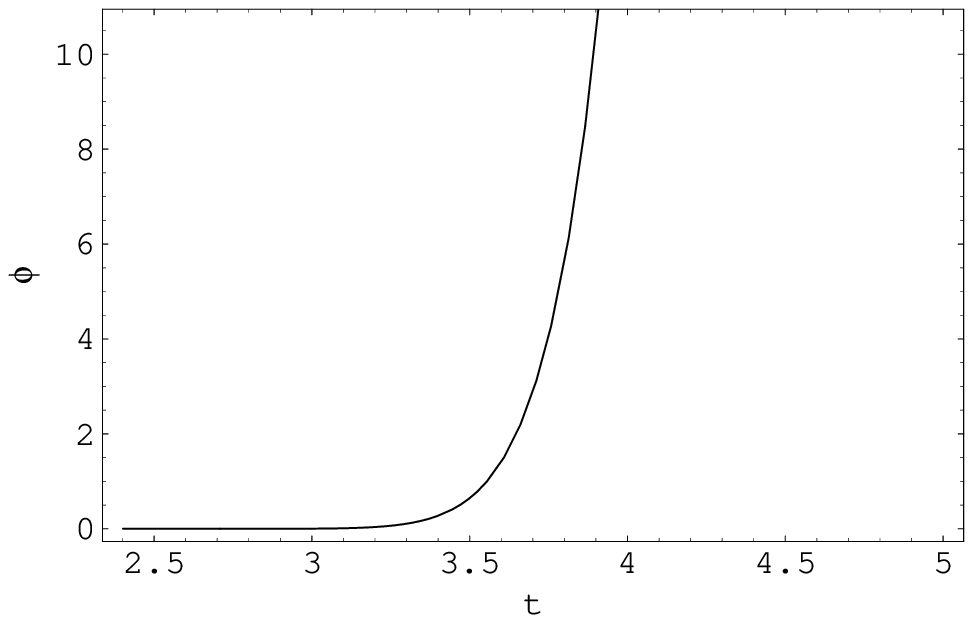}\\
\vspace{1mm} ~~~~~~~~~~~~Fig.2~~~~~~~~~~~~~~~~~~~~~~~~~~~~~~~~~~~~~~~~~~~~~~~~~~~~~~~~~~~~~~~~~~~~~~~~~~~~~~~~~~Fig.3\\

\vspace{7mm}
\end{figure}

\begin{figure}
\includegraphics[height=1.8in]{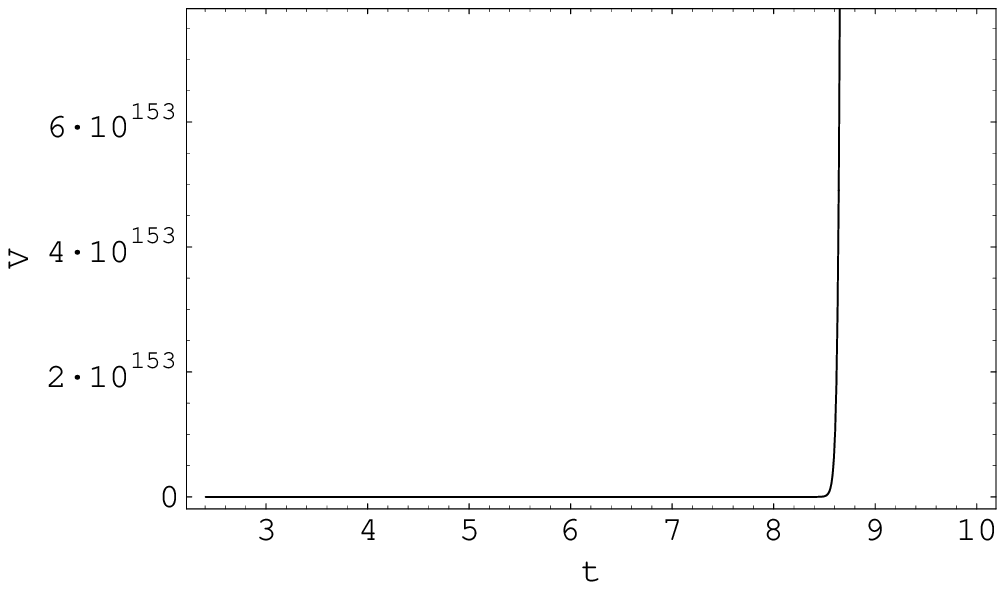}~~~~
\includegraphics[height=1.8in]{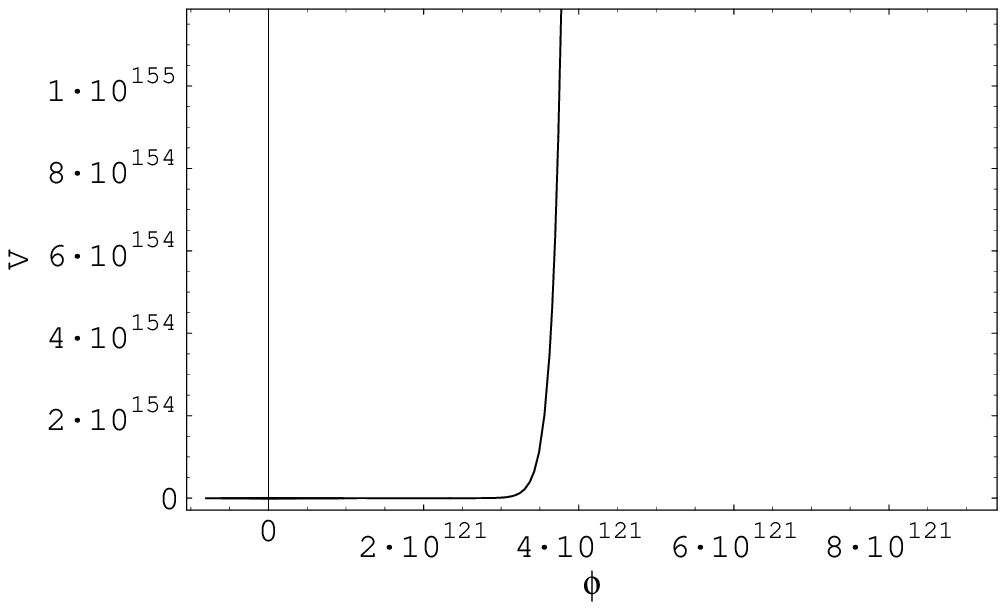}\\
\vspace{1mm} ~~~~~~~~~~~~Fig.4~~~~~~~~~~~~~~~~~~~~~~~~~~~~~~~~~~~~~~~~~~~~~~~~~~~~~~~~~~~~~~~~~~~~~~~~~~~~~~~~~~Fig.5\\

\vspace{7mm}

Figs. 2, 3 and 4 represent the variations of $a$, $\phi$ and $V$
against $t$ respectively and Fig. 5 represents the variation of
$V$ against $\phi$ with
$A=0.1,B=0.1,C=0.001,\delta=0.005,\alpha=1,n=0.6$.

\vspace{6mm}

\end{figure}

In the above expression, $V$ has been expressed as a function of
the scale factor $a$. The variation of $V$ with variation of $a$
is presented in figure 1. It is apparent from the figure 1 that V
has a sharp increase from a lower value with increase in $a$. The
variations of $a$, $\phi$ and $V$ with variation of $t$ has been
presented in figures 2, 3 and 4 respectively. We find that $a$,
$\phi$ and $V$ gradually increases with increase in $t$. In figure
5, we have presented the variation of $V$ with $\phi$ and we see
that $V$ gradually increases with increase in $\phi$. In figure 6,
we have presented the evolution of effective equation of state
parameter $w=\frac{p_{1}+p_{2}}{\rho_{1}+\rho_{2}}$~ with time $t$.
From the graphical representation we have realized the crossing of
$w=-1$ for the interacting dark energy model in evolution of the
universe. So this interacting model generates the whole evolution
of the universe from quintessence to phantom barrier.\\

In this letter, we have considered a flat FRW universe. Instead of
considering only one candidate for dark energy we have considered
phantom field and modified Chaplygin gas. Interaction has been
considered between them. Accordingly, an interaction term has been
introduced in the conservation equation. An expression for the
potential $V$ has been generated under the situation of
interaction. From equation (11), it has been found that the energy
of Chaplygin gas is getting transferred to phantom field.
Subsequently the variation of $V$ has been studied with variation
of the scale factor. It has been observed that the potential of
the phantom field increases from a lower value with evolution of
the Universe. It has been observed that, the field has an
increasing tendency and potential has a increasing tendency with
passage of cosmic time. Thus, it can be concluded that in the
presence of interaction, potential increases and field increases
with evolution of the Universe. We have further realized the
crossing of $w=-1$ for the interacting dark energy model in
evolution of the universe from quintessence to phantom barrier.\\\\\\\\

\begin{figure}
\includegraphics[height=2.5in]{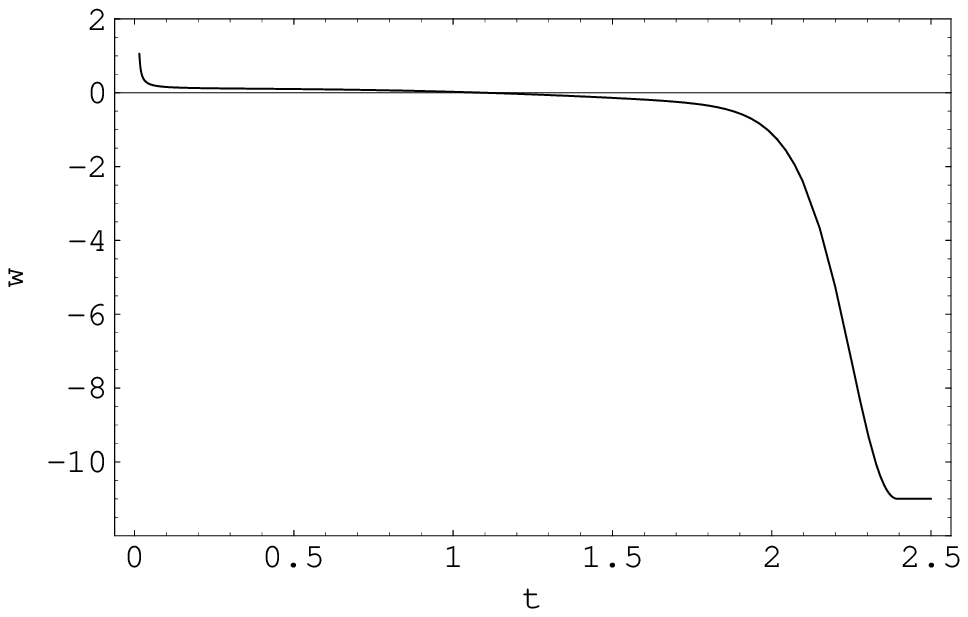}\\
\vspace{1mm} ~~~~~~~~~~~~Fig.6~~~~~~~~~~~~~~~~~\\

\vspace{7mm}

Fig. 6 represents the evolution of $w$ against $t$ for
$A=0.1,B=0.1,C=0.001,\delta=0.005,\alpha=0.5,n=0.6$.

\vspace{6mm}

\end{figure}

{\bf Acknowledgement:}\\

The authors are thankful to IUCAA, India for warm hospitality
where part of the work was carried out. Also UD is thankful to
UGC, Govt. of India for providing research project grant (No. 32-157/2006(SR)).\\

{\bf References:}\\
\\
$[1]$ R. Cai and A. Wang, {\it JCAP} {03} 002 (2005); J. Hao and
X. Li, {\it Phys. Rev. D} {\bf 67} 107303 (2003); P. de Bernardis
et al, {\it Nature} {\bf 404} 955 (2000). N. Bahcall, J. P.
Ostriker, S.
Perlmutter and P. J. Steinhardt, {\it Science} {\bf 284} 1481 (1999).\\
$[2]$ B. Ratra, and P. J. E. Peebles, {\it Phys. Rev. D} {\bf 37}
3406 (1988); J. G. Hao and X. Z. Li, {\it Phys. Rev. D} {\bf 66}
087301 (2002); W. Zimdahl, D. Pavon  and L. P. Chimento, {\it
Phys. Lett. B} {\bf 521} 133 (2001); I. Zlatev, L. Wang and P. J.
Steinhardt, {\it Phys. Rev. Lett.} {\bf 82} 896 (1999);
Z. Guo, Y. Piao, X. Zhang and Y. Zhang, {\it Phys. Lett. B} {\bf 2005} 177 (2005).\\
$[3]$ J. Hao and X. Li, {\it Phys. Lett. B} {\bf 606} 7 (2005);
T. Gonzalez and I. Quiros, {\it Class. Quantum Grav.} {\bf 25} 175019 (2008).\\
$[4]$ Z. Guo and Y. Zhang, {\it Phys. Lett. B} {\bf 645} 326
(2007); U. Debnath, {\it Astrophys. Space. Sci.} {\bf 312} 312
(2007); S. Chattopadhyay and U. Debnath, {\it Grav. Cosmo.} {\bf 14} 341 (2008).\\
$[5]$ A. Vikman, {\it Phys. Rev. D} {\bf 71} 023515 (2005).\\
$[6]$ A. Melchiorri, L. Mersini, C. J. Odmann and M. Trodden, {\it Phys. Rev. D} {\bf 68} 043509 (2003).\\
$[7]$ R. R. Caldwell, {\it Phys. Lett. B} {\bf 545} 23 (2002);
S. M. Carroll, M. Hoffman, and M. Trodden, {\it Phys. Rev. D} {\bf 68} 023509 (2003).\\
$[8]$ WMAP Collab. (D. N. Spergel \emph{et al}), astro-ph/0603449.\\
$[9]$ A. R. Liddle and D. H. Lyth, {\it Cosmological Inflation and Large-Scale Structure} (Cambridge University Press, 2000)\\
$[10]$ E. J. Copeland, M. Sami and S. Tsujikawa, {\it Int. J. Mod. Phys. D} {\bf 15} 1753 (2006)\\
$[11]$ M. R. Setare, {\it Phys. Lett. B} {\bf 644} 99 (2007); {\it
Eur. Phys. J. C} {\bf 52} 689 (2007); Q. G. Huang and M. Li, {\it
JCAP} {\bf 08} 013 (2004); M. R.
Setare, J. Zhang and X. Zhang, {\it JCAP} {\bf 03} 007 (2007).\\
$[12]$ A. Kamenshchik, U. Moschella and V. Pasquier, {\it Phys.
Lett. B} {\bf 511} 265 (2001); V. Gorini, A. Kamenshchik, U.
Moschella and V. Pasquier, gr-qc/0403062.\\
$[13]$ V. Gorini, A. Kamenshchik and U. Moschella, {\it Phys. Rev.
D} {\bf 67} 063509 (2003); U. Alam, V. Sahni, T. D. Saini and A.
A. Starobinsky, {\it Mon. Not. Roy. Astron. Soc.} {\bf 344},
1057 (2003).\\
$[14]$ H. B. Benaoum, {\it hep-th}/0205140; U. Debnath, A.
Banerjee and S. Chakraborty, {\it Class.
Quantum Grav.} {\bf 21} 5609 (2004).\\
$[15]$ M. S. Berger and H. Shojaei, {\it Phys. Rev. D} {\bf 74} 043530 (2006).\\
$[16]$ Z. Guo and Y. Zhang, {\it Phys. Rev. D} {\bf 71} 023501 (2005).\\
$[17]$ P. Singh,  M. Sami and N. Dadhich, {\it Phys. Rev. D} {\bf 68} 023522 (2003).\\
$[18]$ S. Chattopadhyay and U. Debnath, {\it Braz. J. Phys.} {\bf 39} 85 (2009);
 M. Cataldo, P. Mella, P. Minning  and J. Saavedra, {\it Phys. Lett. B} {\bf 662} 314 (2008);
  H. Zhang and Z. H. Zhu, {\it Phys. Rev. D} {\bf 73} 043518 (2006).\\
$[19]$ B. Wang, Y. Gong and E. Abdalla, {\it Phys. Lett. B} {\bf
624} 141 (2005).\\

\end{document}